\documentclass[aps,prl,showpacs,notitlepage,nofootinbib,superscriptaddress,floatfix,showkeys,twocolumn]{revtex4-1}

\usepackage{blindtext}
\usepackage{hyperref}
\usepackage{amsmath,amssymb}
\usepackage{float}
\usepackage{microtype}
\usepackage{graphicx}
\usepackage{bm}
\usepackage{latexsym}
\usepackage{epsfig}
\usepackage{psfrag}
\usepackage{color}
\usepackage[dvipsnames]{xcolor}
\usepackage{subfigure}
\usepackage[section]{placeins}
\def\e1i{\epsilon_{1\mathrm{i}}}

\allowdisplaybreaks[1]

\begin{document}

\title{Constraining dark matter interactions mediated by a light scalar with white dwarfs }

\author{Maura~E.~Ramirez-Quezada}
\email{me.quezada@hep-th.phys.s.u-tokyo.ac.jp}
\affiliation{%
Department of Physics, University of Tokyo, Bunkyo-ku, Tokyo
 113--0033, Japan
}
\affiliation{Dual CP Institute of High Energy Physics, C.P. 28045, Colima, M\'exico}%
\begin{abstract}
Observations of white dwarfs in dark matter-rich environments can provide strong  limits on the strength of  dark matter interactions. Here we apply the recently improved formalism of
the dark matter capture rate
in white dwarfs to  a general model in which dark matter interacts with the white dwarf ion components via a light scalar mediator. We compute the dark matter capture rate in the optically thin limit in a cold white dwarf from the globular cluster Messier. We then estimate the threshold cross-section, which significantly varies as a function of the light scalar mediator mass $m_\phi$ in the range of $0.05\, m_\chi<m_\phi<m_\chi$ and becomes constant when $m_\phi>m_\chi$. We also show that the bounds obtained from the  dark matter capture in a white dwarf from the globular cluster Messier 4  are complementary to direct detection experiments and particularly strong in the sub-GeV regime.
\end{abstract}
\maketitle

\section{Introduction}
The indirect evidence of the existence of dark matter (DM) in the universe is a major milestone in physics. This elusive matter component may interact with Standard Model (SM)  particles in a weak but significant way. The leading candidates are the weakly interacting massive particles (WIMPs)~\cite{Bertone:2004pz}. Typically in direct detection (DD) experiments, the DM particles are expected to interact with  nucleons in the detector by exchanging a marginal momentum. If the interaction  mediator is a massive particle,  it is possible to assume a DM-nucleon point-like interaction. On the other hand, when the interaction is mediated by a light scalar particle, this assumption no longer holds. In this case, the mediator mass is comparable or smaller to the momentum transfer, and the effective field theory (EFT) treatment breaks down~\cite{Fornengo:2011sz,Kaplinghat:2013yxa,Li_2015,Nobile_2015,Kahlhoefer:2017ddj}. 

A promising way to search for DM interactions is to capture DM particles in the core of white dwarfs in DM-rich environments~\cite{Biswas:2022cyh,Amaro-Seoane:2015uny,Panotopoulos:2020kuo,Bell:2021fye}. In particular, white dwarfs (WDs), due to their gravitational potential, are excellent candidates to explore interactions of DM particles with masses in the Sub-GeV mass region~\cite{Bell:2021fye}. The DM particles will fall into the WD gravitational potential and interact with the stellar matter while traversing the star. If the DM loses enough energy after scattering off the stellar material, it will be captured and accumulated in the WD core. The accreted DM within the WD core will thermalize and eventually annihilate into SM particles. The capture, thermalization, and annihilation process will inject energy into the WD and increase its temperature. This extra heating will eventually generate an observable that we can use to set bounds on the interaction strength of the DM particles~\cite{Dasgupta:2019juq,Dasgupta:2020dik}.

In principle, the capture process of DM in WD cores occurs via the scattering of DM with either electrons or ions. In this work, we will apply the recently improved capture rate formalism given in Ref.~\cite{Bell:2021fye} to  a minimal model in which Dirac fermion DM interacts with the WD ion-components via a light scalar mediator $\phi$~\cite{Balaji:2022noj,Kaplinghat:2013xca,Curtin:2013qsa,Cotta:2013jna,Fornengo:2011sz}. Such particles are very well  motivated since their existence have important implications. For instance,  light mediators can lead to long-range attractive forces between DM particles and eventually  enhance the DM annihilation cross-sections at low temperatures via the mechanism of Sommerfeld enhancement~\cite{Hisano:2003ec,Hisano:2004ds,https://doi.org/10.1002/andp.19314030302,Arkani-Hamed:2008hhe,Cassel:2009wt,Liu:2013vha}.

By computing the capture rate of DM in  WDs from the globular cluster Messier 4 (M4) we place limits on the DM-proton interactions in the Sub-GeV mass region.  We show that the bounds obtained by the DM capture are complementary to those of DD experiments. We find that observations of WDs in M4 can set limits in the sub-GeV region of the DM mass for a wide range of the scalar mediator mass. However, it is important to emphasise that  these bounds strongly depend on whether there is DM in the globular cluster M4.

The paper is organized as follows. In Sec. \ref{sec:DM-ion_xsec} we introduce the DM-ion scattering cross section and specify the role of the scalar light mediator. In Sec. \ref{sec:DM_capture} we briefly discuss the capture rate in the optically thin limit and argue why this is a good approximation in our parameters of interest. We present and discuss our results in Sec. \ref{sec:results} and conclude in Sec.\ref{sec:conclusions}.

\section{DM-ion scattering cross section }\label{sec:DM-ion_xsec}
DM particles are expected to have tiny interaction strengths with ordinary matter in many models. For very heavy mediators, the  DM particle interactions with nucleons can be described by effective operators, which are parametrised by the mass scale $\Lambda$~\cite{Fan:2010gt,Fitzpatrick:2012ix},
\begin{equation}
    D_i^N=\frac{c_i}{\Lambda^2}\bar{\chi}\Gamma_i\chi \Bar{N}\Gamma'_iN\label{eq:Di},
\end{equation}
where $c_i$ are the Wilson coefficients, $\Gamma_i$ are Lorentz-invariant combinations of the Dirac matrices and $N=p,n$ for protons and neutrons respectively. Here, we will consider Yukawa interactions with the DM particle, which sets $\Gamma_i,\Gamma_i'=1$. Therefore,  the interaction corresponds to the operator $D_1^N$
\begin{equation}
    D_1^N=\frac{c_1}{\Lambda^2}\bar{\chi}\chi \Bar{N}N\label{eq:D1},
\end{equation}
with 
\begin{equation}
    c_1=\frac{\sqrt{2}m_N}{v}\left[\sum_{q=u,s,d}f_{T_q}^N+\frac{2}{9}f^N_{TG}\right]\label{eq:c1},
\end{equation}
where $v=246\,\rm GeV$ is the electroweak (EW) vacuum expectation value of the Higgs field, $m_N$ is the nucleon mass  and $f_{T_q}$ and $f_{T_G}$ are the hadronic elements~\cite{Ellis:2018dmb,Belanger:2013oya}. The expression in Eq.~\eqref{eq:c1} only holds for the particular case when the couplings of the mediator with quarks are proportional to their masses (Yukawa couplings). This situation is realised if the mediator couples with the Standard Model sector only via mixing with the Higgs boson;
A more generic expression for different types of interactions is given in, e.g., Eq.(C.6) in Ref.~\cite{Fan:2010gt}.

Due to the very low energies involved in DM-nucleon scattering, such interaction shall be  described at the nucleus level by non-relativistic operators $D_i^{NR}$. This is obtained by contracting the fermionic DM and nucleon bilinears. For the bilinears in Eq.~\eqref{eq:D1}, the  non-relativistic  expansion corresponds to~\cite{Cirelli:2013ufw},
\begin{equation}
    \bar{\chi}\chi \Bar{N}N=4m_\chi m_N D_1^{NR},\label{eq:DNR}
\end{equation}
where $m_\chi$ is the DM mass and $D_1^{NR}=1$.

Here we will consider a light Dirac fermion DM candidate coupling to the SM via a light scalar $\phi$~\cite{Kaplinghat:2013xca,Kaplinghat:2013yxa}.  Since this mediator is light the effective operator approach is modified by the simple replacement $\Lambda^2\to(q^2+m_\phi^2)/(g_\chi g_{SM})$, where $q$ is the transferred momentum in the $t$-channel, $m_\phi$ is the light scalar mediator mass and $g_{\chi,SM}$ are the couplings of the scalar to the DM and the SM particles respectively.
Therefore, for the Yukawa interaction, the DM-nucleon amplitude is given as
\begin{equation}
    \mathcal{M}_N=4 g_\chi g_{SM} c_1 \frac{ m_\chi m_N}{q^2+m_\phi^2}.
\end{equation}

From the EFT approach, we then notice that spin-averaged squared amplitude for scattering off a target nucleus $T$ with mass $m_T$ is given by
\begin{equation}
    |\mathcal{M}_T|^2=\frac{m_T^2}{m_N^2}\langle|\mathcal{M}_N|^2 F(E_R)^2\rangle,\label{av_eneg}
\end{equation}
 and $F(E_R)^2$ is  commonly used Helm form factor~\cite{PhysRev.104.1466,Duda_2007} and $E_R=q^2/2m_T$ is the recoil energy. Here we have averaged over the energy; specifically, we adopt the following form~\cite{LEWIN199687}, 
\begin{equation}
 \langle|\mathcal{M}_N|^2  F(E_R)^2	\rangle=   \!\!\!\int_0^{E_R^{\rm max}} \!\!\!\!\!\!|\mathcal{M}_N|^2 \left(\frac{3j_1(qR_1))}{q R_1}\right)^2\frac{e^{-q^2s^2}}{E_R^{\rm max}} dE_R,
\end{equation}
here $j_1$ is the spherical Bessel function of the first
kind, $R_1$ is the effective nuclear radius, and $s$ is the nuclear skin thickness, see Ref.~\cite{LEWIN199687}. $E_R^{\rm max}$ is the maximum recoil energy found when the momentum transfer $q$ is maximised, $E_R^{\rm max}=q^{2\,\rm max}/2m_T$.
The differential scattering cross-section in the non-relativistic approximation is given by
\begin{equation}
    \frac{d\sigma}{d\cos\theta}=\frac{1}{32\pi}\frac{ |\mathcal{M}_T|^2}{(m_\chi+m_T)^2}.
    \end{equation}

    In the centre-of-mass (cm) frame, we express the momentum transferred in terms of the cm angle $\theta$~\cite{Busoni:2017mhe}
    \begin{equation}
       q^2=4\sin^2(\theta/2)v_r^2\frac{m_\chi^2 m_T^2}{(m_T+m_\chi)^2}.
    \end{equation}
where $v_r$ is the relative velocity between the DM and the target.

    Putting all of this together, the differential scattering cross-section for a Yukawa DM-ion interaction is given as
   \begin{equation}
          \frac{d\sigma}{d\cos\theta}=\frac{c_1^2 g_\chi^2 g_{SM}^2\beta^2}{2\pi}\left\langle \frac{F(E_R)^2}{(2m_T E_R+m_\phi^2)^2}\right\rangle,\label{eq:DM-ion_cross-section}  \end{equation}
where $\beta$ is the reduced mass of the system.
Notice that if in Eq.~\eqref{eq:DM-ion_cross-section} we define $\alpha^2=g_\chi^2 g_{SM}^2c_1^2/(8\pi)$ we can recover the expression given in Ref.~\cite{Dasgupta:2020dik}.

\section{Dark matter capture in white dwarfs}\label{sec:DM_capture}
\begin{figure}
    \centering
    \includegraphics[width=0.5\textwidth]{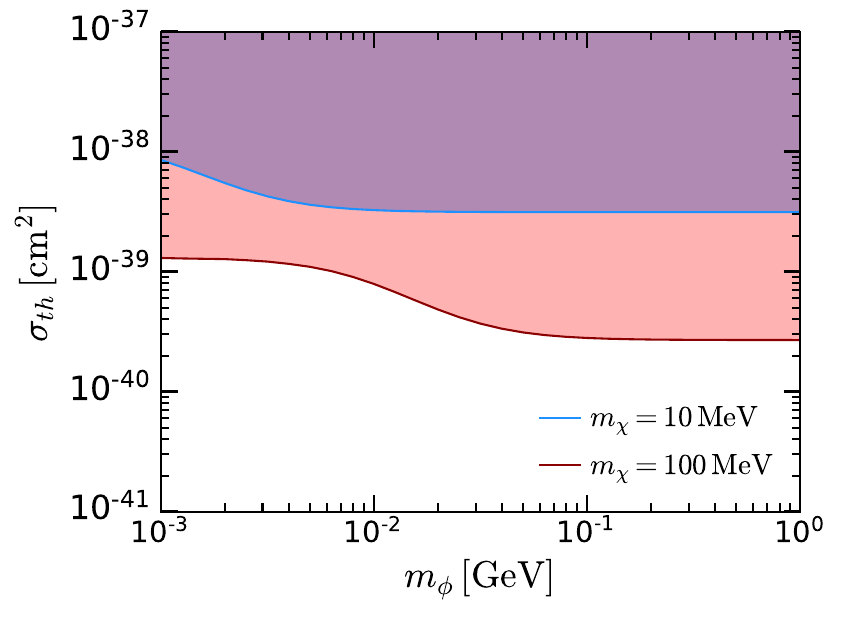}
    \caption{Threshold cross-section for $m_\chi = 10\,\rm MeV$ (blue) and $m_\chi= 100\,\rm MeV$ (brown)  as a function of the light mediator mass $m_\phi$. The shaded area in both cases indicates the interactions in which  the capture process operates maximally.}
    \label{fig:xsec_th}
\end{figure}
The capture of DM in WDs is possible via the scattering process with electrons and ions~\cite{Bell:2021fye}. Since in the general model studied here,  DM interacts with the nucleon via a scalar  mediator~\cite{Kaplinghat:2013xca,Kaplinghat:2013yxa}, we will account only for the latter process. Therefore, 
the capture rate  of DM in WDs by ions in the optically thin limit is given by~\cite{Bell:2021fye}
\begin{equation}
    C=\frac{\rho_\chi}{m_\chi}\int_0^{R\star}dr4\pi r^2\int_0^\infty du_\chi \frac{w}{u_\chi}f_{\rm MB}(u_\chi)\Omega^-(w),\label{eq:Cthin}
\end{equation}
where $R_\star$ is the WD radius, $\rho_\chi$ is the DM density, $u_\chi$ is the DM velocity at infinity, and $w$ is the DM velocity after it falls close to the WD with $w^2=u_\chi^2+v_e(r)^2$. $f_{\rm MB}$ is the Maxwell-Boltzmann DM distribution function, which in the zero temperature limit $T_\star\to0$ is given by~\cite{Busoni:2017mhe,Bell:2021fye}
\begin{equation}
f_{\rm MB}(u_\chi)=\frac{u_\chi}{v_dv_\star}\sqrt{\frac{3}{2\pi}}
\left[e^{-\frac{3}{2v_d^2}\left(u_\chi-v_\star\right)^2}-e^{-\frac{3}{2v_d^2}\left(u_\chi+v_\star\right)^2}\right].\label{eq:distribution_relative_distribution}
\end{equation}
where $v_\star$ is the WD velocity in the globular cluster rest frame and $v_d$ is the velocity dispersion of the DM. 

Finally, $\Omega^-(w)$ is the interaction rate probability given as a function of the differential scattering cross-section and the ion number density. For ion targets in a WD  in  the zero temperature limit, it is given by~\cite{Busoni:2017mhe}
\begin{eqnarray}
\Omega^-(w)&=& 
\frac{4\mu_+^2}{\mu w}n_T(r) \int_{w\frac{|\mu_-|}{\mu_+}}^{v_e}dv v \dfrac{d\sigma}{d\cos \theta}.\label{eq:capture_prob_specific}
\end{eqnarray}
where  $\mu=m_\chi/m_T$, $\mu_\pm=(\mu\pm1)/2$. The capture process happens when the DM velocity after scattering $v$ is below the escape velocity $v_{e}$ of the WD. 

The case where all DM traversing the star is captured, i.e.  the geometric limit, occurs when $\Omega^-(w)\to 1$. In this case, we can write the capture rate as~\cite{Bell:2021fye},
 \begin{align}
 C_{geom}=\frac{\pi R_\star^2 \rho_\chi}{3v_\star m_\chi} \Bigg[(3 v_{e}^2(R_\star)+&3 v_\star^2+v_d^2) \mathrm{erf} \left(\sqrt{\frac{3}{2}}\frac{v_\star}{v_d}\right) \nonumber\\[0.2cm]
& \left.+\sqrt{\frac{6}{\pi}} v_\star v_d e^{-\frac{3v_\star^2}{2 v_d^2}}\right]. \label{eq:CWD_geo_limit}
\end{align}
This limit is a very good approximation for DM-cross sections above a threshold cross-section $\sigma_{th}$. The threshold cross-section is where the geometric and optically thin limits intersect.\footnote{For DM scattering cross-sections of $\sigma\sim\sigma_{th}$ we shall introduce the star opacity as discussed in Ref.\cite{Bell:2021fye}.}
However,  the DM-ion cross-sections we will study are below this threshold, and it is safe to consider the optically thin limit given in Eq. \eqref{eq:Cthin}.

In Fig.~\ref{fig:xsec_th}, we show the threshold cross section for two different DM masses $m_\chi= 10\,\rm MeV$ (blue) and $m_\chi =100\,\rm MeV$ (red). The threshold cross-section varies accordingly to the light mediator mass and the $g_\chi g_{g_{SM}}$ parameter, which is not shown in the plot. When the light mediator mass is large ($m_\phi\to\infty$), the threshold cross-section becomes independent of $m_\phi$, and we recover  the EFT operator approach. This situation is reached once the light mediator mass reaches the DM mass, and hence it happens first for the case of $m_\chi=10\,\rm MeV$ as shown in the  blue line in Fig.~\ref{fig:xsec_th}. On the other hand, if the light mediator mass is too small ($m_\phi<<m_\chi$), the DM mass dominates in the expression of Eq.~\ref{eq:DM-ion_cross-section} and hence the threshold cross-section becomes independent of the light mediator mass. The mass interval at which the threshold cross-section is sensitive to the scalar mass is $0.05\, m_\chi<m_\phi<m_\chi$ The shaded area indicates that all cross-sections in this region belong to the geometric limit, and DM is maximally captured in the WD.

We must compute the capture rate for DM cross-sections below the threshold cross-sections to obtain reliable bounds on DM-proton interaction. 

\section{Results and discussion}\label{sec:results}
\begin{figure*}
    \centering
\includegraphics[width=0.49\textwidth]{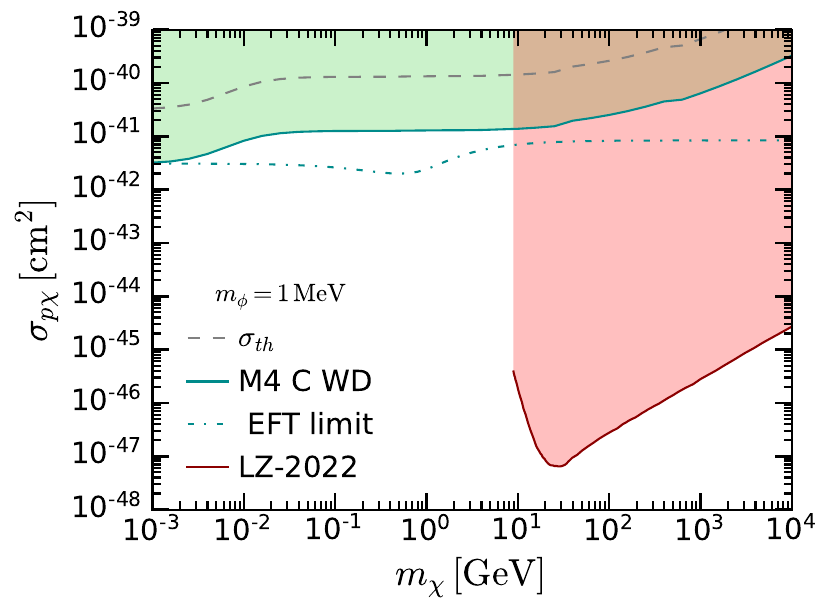}
    \includegraphics[width=0.49\textwidth]{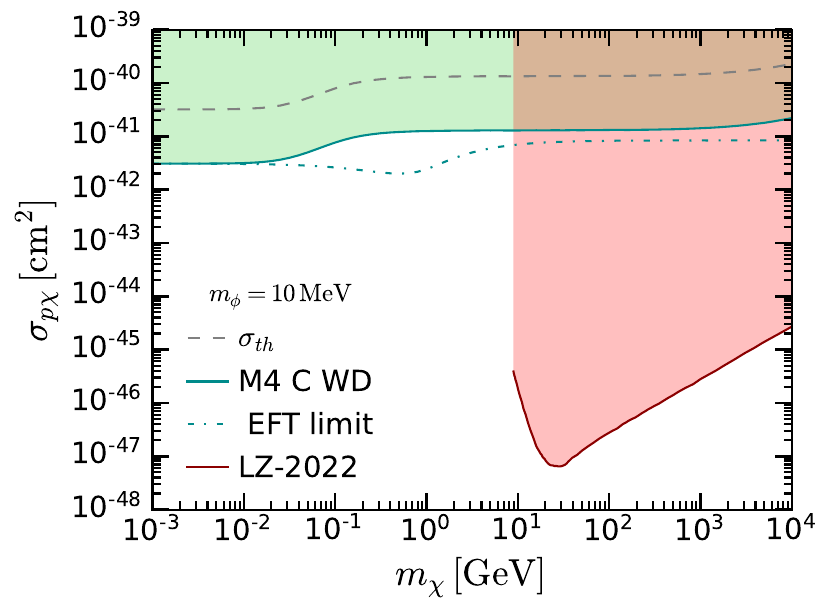}
    \caption{DM-proton interactions induced by scalar light mediator $\phi$ with a mass of $m_\phi=1\,\rm MeV$ (left panel) and $m_\phi=10\,\rm MeV$ (right panel). The bounds shown in green have been computed assuming a DM density of $\rho_\chi=798\,\rm GeV\,cm^{-3}$ in the globular cluster M4~\cite{McCullough:2010ai}. We show in red the latest constraints set by LZ for comparison~\cite{LZ:2022ufs}. The dashed-dotted line represents the EFT limit case of the DM-proton cross-section.}
    \label{fig:DM_bounds}
\end{figure*}
Hubble Space Telescope (HST) observations of cold WDs have been used to determine the age of M4~\cite{Hansen:2004ih,2009ApJ...697..965B}. This data was later translated into luminosity and effective temperatures in Ref.~\cite{McCullough:2010ai}, which allows us to  deduce the WD radii and their related masses assuming a mass-radius relation. To compute the DM capture rate, we have used the coldest  WD observed  in the globular cluster M4. 
We consider a WD made entirely of Carbon-12.
Using the Salpeter equation of state~\cite{1961ApJ134669S,1961ApJ134683H} coupled with the Tolman--Oppenheimer--Volkoff (TOV) equation, we model the inner structure of the coldest M4 WD corresponding to a WD with a mass of $M_\star=1.38\,\rm M_\odot$ and a radius of $R_\star=8.7\times10^2\,\rm km$.  We obtain the radial profiles for the target number density $n_T(r)$ and the WD escape velocity $v_e(r)$.  It is worth noting that the radius reported in this work slightly differs from that reported in Ref.\cite{Bell:2021fye} where a different equation of state was used~\cite{PhysRev.75.1561,Rotondo:2011zz}. However, the difference is only 1.4 times the radius reported in this work.

If DM is present in M4,
the expected DM density is very large when compared to the DM  in the
Solar vicinity, which enhances the capture rate. Using a Navarro-Frenk-White profile to model the DM halo, the DM density in M4 is expected to be as large as $\rho_\chi=798\,\rm GeVcm^{-3}$~\cite{McCullough:2010ai}. The exact values of the star and DM dispersion velocities for WDs in M4 are unknown. We assume the star velocity  to be $v_\star=20\,\rm km\,s^{-1}$ while the dispersion velocity is found to be less than $v_d=8\,\rm km\,s^{-1}$~\cite{McCullough:2010ai}.

This allows us to compute the capture rate in the optically thin limit given in Eq.~\eqref{eq:Cthin}. Nonetheless, we should not exceed the threshold cross-section $\sigma_{th}$ to compute the DM-ion interaction limits using the capture rate in the thin limit. Furthermore, for the DM mass range  studied here, the capture process is successful after a single scattering~\cite{Bell:2021fye}.\footnote{This is true for DM masses below a threshold DM mass which in the case of a WD of $M_\star=1.38\,\rm M_\odot$  the threshold mass is $m_{th}\sim 10^4$ GeV~\cite{Bell:2021fye}.}

The DM capture and annihilation rates are already in equilibrium in the oldest WDs~\cite{Bertone:2007ae}. This means that the DM annihilation rate is equal to the capture rate $C$ given in Eq.~\eqref{eq:Cthin}. Hence, the heating luminosity observed at a distance is estimated to be,
\begin{equation}
    L_{\chi}=m_\chi C. \label{eq:DM_luminoity}
\end{equation}
We compare the WD observed luminosity $L_\gamma\sim10^{31}\,\rm GeV\,s^{-1}$ to the fraction luminosity due to the DM in the WD core. By requiring that the luminosity $L_\gamma$ be larger than $L_\chi$, we constrain the DM-proton cross-section.  

In Fig.~\ref{fig:DM_bounds}, we show the upper bounds for the DM-proton cross-section, induced by a light mediator, given by the M4 carbon WD (green line). To compute these bounds, we have assumed that all the observed luminosity comes from the DM effects, i.e. $L_\gamma=L_\chi$. 

In the left panel, we show the DM-proton cross-section bounds for a light mediator of 
mass $m_\phi=1\,\rm MeV$ while in the right panel, the mass is $m_\phi=10\,\rm MeV$. 
For comparison, we also show the bounds from the latest results given by LUX-ZEPLIN (LZ)~\cite{LZ:2022ufs}.
The areas shaded in green, and red corresponds to the excluded parameter space by WD in M4 and LZ, respectively.   We observe that the limit given by LZ is stronger than that provided by the M4 WD for the two possible mediator masses in the DM mass region above a  few GeV.  The dashed grey line corresponds to the threshold cross-section. 

Additionally, we have checked that the couplings are always perturbative in the region of interest. 
The product $g_\chi g_{SM}$ that fulfils $L_\gamma=L_\chi$ remains small ($g_\chi g_{SM}\lesssim \mathcal{O}(1) )$ as long as $m_\phi\lesssim 
m_\chi$.
In dashed-dotted lines, we have also added the WD bounds that would be given in the EFT limit  of the DM-proton cross-section. We observe that in both cases, the limits given by WDs using this approach are overestimated and become a good approximation when $m_\chi\lesssim m_\phi$. The EFT approach agrees with the one  reported in \cite{Dasgupta:2019juq} in the low-mass regime, and there is a noticeable difference in the large-mass regime. This is due to a different choice of WD in M4 with a mass of $0.4\, \rm M_\odot$ and a radius of $9000$ km. This difference was also found in Ref.~\cite{Bell:2021fye}, which agrees with our results by a factor of $\sim 3$ in the low-mass regime and  up to a factor of $\sim 9$ in the large-mass regime due to the choice of EOS.

 On the other hand, since the DM evaporation mass in WDs is of the order of keV, it is possible to constrain the interactions for Sub-GeV DM~\cite{Bell:2021fye}. This means that the bounds given by the WD are valid as long as the DM mass is above this evaporation mass. Otherwise,  the evaporation process will reduce the number of accumulated DM particles in the WD core. Notice that choices  of scalar masses below 1 MeV require lower DM masses, and limits are no longer reliable.

 Here, we have shown that if DM is present in the core of M4, it is clear that the bounds to DM interactions via a scalar mediator provided by WDs are complementary to direct detection experiments for a wide range of scalar mediator masses. If the DM density is a few $\rm GeV/cm^3$ as estimated in Ref.~\cite{Hooper:2010es}, instead of $\simeq 10^2$-$10^3\,\rm GeV/cm^3$  the WD limits  become several orders of magnitude weaker. Finally, it is worth mentioning that such states are actively searched in collider experiments, and  constraints have been given for scalar and pseudoscalar mediators above 10 GeV~\cite{ATLAS:2021hza,ATLAS:2022ygn}. However, searching at colliders might be complicated for very light mediators since these are long-lived particles  that might decay outside the detectors.

\section{Conclusion}\label{sec:conclusions}
In this work, we studied a general model for DM interactions with nuclei induced by light mediators. 
We have discussed the DM capture in WDs and have computed the threshold cross-section as a function of the light mediator mass for two choices of DM mass. The threshold cross-section $\sigma_{th}$ varies according to the scalar light mediator mass  and coupling when $m_\phi<m_\chi$. However, when the light mediator mass reaches the DM mass, this dependence disappears, and the threshold cross section becomes constant for all masses $m_\phi>m_\chi$. We also observed that when the DM mass is much larger than the light mediator $m_\chi>>m_\phi$, the DM mass dominates in the threshold cross-section expression, and hence it becomes independent of the light mediator mass and  only varies accordingly to the DM mass. The threshold cross-section is sensitive to the scalar mass in the interval $0.05\, m_\chi<m_\phi<m_\chi$

Using WDs from the globular cluster M4, we set limits on these interactions accounting for different light mediator masses $m_\phi=1\rm MeV$ and $m_\phi=10\rm MeV$.
We found that if DM is present in M4, the bounds set by WDs  on DM interactions with the nucleus  via the exchange of a scalar light mediator are complementary to those given by direct detection experiments. Specifically, we can constrain interactions for DM masses in the sub-GeV regime for a wide range of $m_\phi$.


\color{black}
\section{Acknowledgements}
The Author would like to thank Shyam Balaji, 
Koichi Hamaguchi, Natsumi Nagata and Yongchao Zhang for
helpful discussions and feedback on the draft. 
MRQ is supported  by the JSPS KAKENHI Grant Number 20H01897.


\bibliographystyle{apsrev4-1}
\bibliography{references.bib}

\end{document}